\documentstyle[12pt,fullpage,subeqn]{article}
\def\be{\begin{equation}}
\def\ee{\end{equation}}
\def\beq{\begin{eqnarray}}
\def\eeq{\end{eqnarray}}

\def\gsim{\:\raisebox{-0.5ex}{$\stackrel{\textstyle>}{\sim}$}\:} 

\begin{document}
\begin{flushright}
MADPH-96-944 \\ IISc-CTS-11/96 \\ TIFR/TH/96-25 \\ hep-ph/9605447
\end{flushright}
\bigskip
\begin{center}
{\Large{\bf SUSY CONTRIBUTIONS TO $R_b$ AND TOP QUARK DECAY}} \\[1cm]
{\large Manuel Drees$^1$, R.M. Godbole$^2$, Monoranjan Guchait$^3$, \\ 
Sreerup Raychaudhuri$^3$ and D.P. Roy$^3$}
\end{center}
\bigskip
\begin{enumerate}
\item[{}] $^1$Physics Dept., University of Wisconsin, Madison, WI
53706, USA.
\item[{}] $^2$Centre for Theoretical Studies, Indian Inst. of Science,
Bangalore 560 012, India (on leave of absence from the Department of 
Physics, University of Bombay, Mumbai, India.

\item[{}] $^3$Theoretical Physics Group, Tata Inst. of Fundamental
Research, \\ Mumbai 400 005, India.
\end{enumerate}
\bigskip\bigskip\bigskip

\begin{center}
{\large{\bf Abstract}}
\end{center}

Stop contributions to radiative corrections to $R_b$ and the top quark 
decay are analysed over the relevant MSSM parameter space. One sees a 
30\% increase in the
former along with a similar drop in the latter in going from the
higgsino dominated to the mixed region.  Consequently one can get a
viable SUSY contribution to $R_b$ within the constraint of the top
quark data only in the mixed region, corresponding to a photino
dominated LSP.  We discuss the phenomenological implications of this
model for top quark decay and direct stop production, which can be
tested with the Tevatron data. \\

\vskip 10pt

Pacs Nos: 12.15.Lk, 14.65.Ha, 14.80.Ly 

\newpage

\noindent {\large{\bf 1. \underbar{\bf INTRODUCTION}}}
\medskip

One of the most intriguing results from the precision measurement of
$Z$ boson parameters at LEP is the $R_b$ anomaly i.e. the ratio
\be
R_b = \Gamma^{\bar bb}_Z/\Gamma^{\rm had}_Z
\ee
is observed to be about $3\sigma$ higher than the standard model (SM)
prediction.  This has aroused a good deal of theoretical interest for
two reasons.  Firstly, there is a natural source for a significant
contribution to this quantity from the minimal supersymmetric extension of the
standard model (MSSM) \cite{one}, due to the large top quark mass.
Secondly, such a contribution would reduce the SM contribution to
$\Gamma^{\rm had}_Z$ slightly and bring the resulting $\alpha_s (M_Z)$
in better agreement with its global average value \cite{two} of
\be
\alpha_s (M_Z) = .117 \pm .005.
\ee
It should be mentioned here that the measured value of $R_c$ seems to
be $1.7\sigma$ below the SM prediction. But there is no natural theoretical
source for this deficit. One can accommodate this by invoking extra
fermions \cite{three} or an extra $Z$ boson \cite{four}. But then one
has to assume an exact cancellation between their contributions to
$R_q \ (q = u, d, s, c, b)$ in order to preserve the agreement of the
extremely precise measurement of $\Gamma_z^{\rm had}$ with its SM
prediction. Thus it is fair to surmise that the $R_c$ anomaly does not
have the same experimental or theoretical significance as $R_b$.
Following the standard practice, we shall explore the $R_b$ anomaly by
assuming $R_c$ to be equal to its SM value of 0.172. With this
assumption, the current experimental value of $R_b$ is \cite{five}
\be
R_b^{\rm exp} = 0.2202 \pm 0.0016 , 
\ee
which is 2.8$\sigma$ above the SM value of 
\be
R_b^{\rm SM} = 0.2157 (0.2158) \ \ {\rm for} \ m_t = 175 (170) GeV .
\ee

There are two MSSM solutions to the $R_b$ anomaly corresponding to the
two distinct regions, $\tan \beta \simeq 1$ and $\sim m_t/m_b$, where
$\tan \beta$ is the ratio of the two higgs vacuum expectation values.
The relevant MSSM contribution comes from the radiative correction
involving stop--chargino exchanges in the first case, while the
dominant contribution comes from the higgs exchange in the second case
\cite{six, seven, eight}. Correspondingly one expects a significant
contribution to  top quark decay from the stop--neutralino and charged
higgs channels respectively. In the present work we shall be
concentrating in the first case, i.e., $\tan \beta \simeq 1$. 

Admittedly there is a vast literature analysing the MSSM contribution
to $R_b$ in the low $\tan \beta$ region \cite{seven}. However, there is as yet
no systematic exploration of the MSSM parameter space to obtain the
best solution to the $R_b$ anomaly, while taking account of the
constraint from top quark decay simultaneously. The present work is
devoted to this exercise. In particular we shall see that, contrary to
the popular notion, there is no viable solution to the $R_b$ anomaly
from the higgsino dominated region, once the top decay constraint is
taken into account. With this constraint, by far the best solution
comes from the mixed region, corresponding to a photino dominated LSP. 

In the following section we briefly discuss the MSSM formalism along
with the relevant formulae for SUSY contributions to $R_b$ as well as
top quark decay. In the next section we shall present our results for
SUSY contributions to $R_b$ and the top branching ratio over a wide
range of the MSSM parameters and identify the region that gives the
best solution to the $R_b$ anomaly within the constraints from top
quark decay. We shall also discuss the phenomenological implication of
this model for top quark decay and direct stop production, which can
be tested with Tevatron data. We shall conclude with a summary of our
results. 

\vspace{.5in}

\noindent{\large{\bf 2. \underbar{\bf FORMALISM} :}}
\medskip

If squarks are degenerate at the Planck or GUT scale, 
the large top quark mass implies the following mass hierarchy among
the right and left handed stops and the remaining squarks at the weak 
scale, 
\be
m_{\tilde t_R} < m_{\tilde t_L} < m_{\tilde q} .
\ee
After mixing the lighter stop
\be
\tilde t_1 = \cos \theta_{\tilde t} \tilde t_R - \sin \theta_{\tilde
t} \tilde t_L
\ee
can have a significantly smaller mass than the other squarks. We shall
be primarily interested in this stop, which is expected to have a
dominant $\tilde t_R$ component. 

We assume that the soft masses of the 
SU(2) $\times$ U(1) $\times$ SU(3) gauginos are
related via the GUT relations, 
\be
M_1 = {5 \over 3} \tan^2 \theta_W \ M_2 \simeq 0.5 M_2 .
\ee
\be
M_3 = {\alpha_S \over \alpha} \sin^2 \theta_W M_2 \simeq 3.5 M_2.
\ee
Thus all the gaugino masses are given in terms of a single mass
parameter $M_2$, while the higgsino masses are controlled by the
supersymmetric mass parameter $\mu$ \cite{one}. The SU(2) and U(1)
gauginos mix with the two higgsino to form the physical neutralino
($\tilde Z_i$) and chargino ($\tilde W_i$) states, i.e.,
\be
\tilde Z_i = N_{i1} \tilde B + N_{i2} \tilde W^3 + N_{i3} \tilde H_1^0
+ N_{i4} \tilde H_2^0 , 
\ee
\be
\tilde W_{iL} = V_{i1} \tilde W_L^\pm + V_{i2} \tilde H_L^\pm , \tilde
W_{iR} = U_{i1} \tilde W_R^\pm + U_{i2} \tilde H^\pm_R .
\ee
The masses and compositions of the chargino and neutralino states are
determined by the three MSSM parameters -- $M_2, \mu$ and $\tan \beta$.
The lightest neutralino $\tilde Z_1$ is assumed to be the lightest
superparticle (LSP). 

The SUSY contribution to $R_b$ can be written as \cite{six}
\be
\delta R_b = R_b^{\rm SM} (0) \left[ 1 - R_b^{\rm SM} (0) \right] \left[
\nabla^{\rm SUSY}_b (m_t) - \nabla^{\rm SUSY}_b (0) \right] ,
\ee
where $R_b^{\rm SM} (0)$ = 0.2196 represents the SM value at $m_t = 0$. 
\[
\nabla_b^{\rm SUSY} (m_t) = {\alpha \over 2\pi \sin^2 \theta_W} \cdot
{v_L F_L + v_R F_R \over v_L^2 + v_R^2}, 
\]
\be
v_L = - {1 \over 2} + {1 \over 3} \sin^2 \theta_W \ , \ v_R = {1
\over 3} \sin^2 \theta_W 
\ee
The SUSY contributions to $Z \rightarrow \bar b b$ come from the
triangle diagrams involving $\tilde W_i \tilde W_j \tilde t_k$ and
$\tilde t_i \tilde t_j \tilde W_k$ exchanges as well as the $\tilde
t_i \tilde W_j$ loop insertions in the $b$ and $\bar b$ legs. The
relevant formulae can be found in \cite{six}. We shall only state them
for the $\tilde W_i \tilde W_j
\tilde t_k$ contribution, in a form more convenient for our discussion. 
\beq
F_{L, R} &=& \sum_{i, j, k} \Biggl[\ O^{L,R}_{ij} M_{\tilde W_i}
M_{\tilde W_j} C_0  \nonumber \\ [3mm]
&& + O^{R,L}_{ij} \left\{ - M^2_Z (C_{23} + C_{12}) - {1 \over 2} + 2
C_{2 4} \right\} \Biggr] \Lambda^{L,R}_{ki} \Lambda^{*L,R}_{kj} 
\eeq
\be
\Lambda^L_{1i} = V^\ast_{i1} \sin \theta_{\tilde t} 
- {m_t \over \sqrt{2} M_W \sin
\beta} V^\ast_{i2} \cos \theta_{\tilde t} , \ \Lambda^R_{1i} = {- m_b
\over \sqrt{2} M_W \cos \beta} U_{i2} \sin \theta_{\tilde t} ,
\ee
\be
O^L_{ij} = - {1 \over 2} \left[ \cos 2 \theta_W \delta_{ij} +
U^\ast_{i1} U_{j1} \right] , \ O^R_{ij} = - {1 \over 2} \left[
\cos 2 \theta_W \delta_{ij} + V^\ast_{i1} V_{j1} \right] , 
\ee
where the $C$ functions are the conventional Passarino--Veltman
functions with arguments ($M_{\tilde W_i} , m_{\tilde t_k} , M_{\tilde
W_j}$) \cite{nine}. The $\Lambda^{L,R}_{1i}$ represent the $b \tilde
t_1 \tilde W_i$ couplings, which are common to the other SUSY
diagrams. The dominant contribution to (14) comes from the $b_L \tilde
t_{1R} \tilde W_1$ Yukawa coupling which favours low $\tan \beta
(\simeq 1)$ and large $V_{12}$ -- i.e. the higgsino dominated region.
On the other hand $O^{L,R}_{ij}$ represent the $Z \tilde W_i \tilde
W_j$ couplings. The analogous factor for the $\tilde t_1 \tilde t_1
\tilde W_k$ contributions corresponds to the U(1) coupling of $Z$ to
$\tilde t_{1R}$, which is relatively small ($\sim \sin^2 \theta_W$).
It is evident from Eq. (15) that large $O^{L,R}_{11}$ favour large
$U_{11} , V_{11}$ -- i.e. large gauge components of $\tilde W_1$
\cite{eight}. Thus the combined requirements of large $\Lambda$ and
$O$ couplings favour a $\tilde W_1$ having large higgsino component in
$V (V_{12})$ and gaugino component in $U (U_{11})$ and/or comparable
$\tilde W_1$ and $\tilde W_2$ masses. As pointed out in \cite{eight},
these conditions cannot be satisfied for $\mu > 0$. Consequently the
best values of $\delta R_b$ for positive $\mu$ are about half of those
for negative $\mu$. Therefore we shall concentrate on the latter case.
In this case the above conditions favour the mixed region ($| \mu |
\sim M_2$) over the higgsino dominated one $(| \mu | \ll M_2)$. 
Indeed we shall see that one gets typically 30\% larger values of
$\delta R_b$ in the former region compared to the latter. 

One has to assume $\tilde W_1 , \tilde t_1$ masses as well as $\tan
\beta$ close to their lower limits in order to obtain significant
values of $\delta R_b$. Under these assumptions one predicts a
significant SUSY contribution to top decay from $t \rightarrow \tilde
t_1 \tilde Z_i$. The relevant formalism has been discussed in
\cite{eight, ten}. We shall only state the final result. 
\beq
\Gamma (t \rightarrow \tilde t_1 \tilde Z_i) &=& {\alpha m_t \over 16
\sin^2 \theta_W} \sqrt{1 - 2 (x + y_i) + (x - y_i)^2} \nonumber \\ [3mm]
&& \left[ \left( | C_L^i |^2 + | C_R^i |^2 \right) (1 - x + y_i) + 4
\sigma_i Re \left(C_L^{i\ast} C_R^i\right) \sqrt{y_i} \right] ,
\eeq
where $x = m^2_{\tilde t_1} / m^2_t \;, \ y_i = M^2_{\tilde Z_i} /
m^2_t$ , $\sigma_i = sgn(M_{\tilde Z_i})$ and 
\beq
C_L^i &=& \left( {1 \over 3} \tan \theta_W N_{i1} + N_{i2} \right)
\sin \theta_{\tilde t} + {m_t \over M_W \sin \beta} N_{i4} \cos
\theta_{\tilde t} \ , \nonumber \\ [3mm]
C_R^i &=& - {4 \over 3} \tan \theta_W N_{i1} \cos \theta_{\tilde t} +
{m_t \over M_W \sin \beta} N_{i4} \sin \theta_{\tilde t} ,
\eeq
represent the $t \tilde t_1 \tilde Z_i$ couplings (compare Eq. 14).
The dominant contributions come from the $t_L \tilde t_{1R} \tilde
N_i$ and $t_R \tilde t_{1L} \tilde N_i$ Yukawa couplings represented
by the last terms in $C^i_{L, R}$. Thus the favoured decay channel
corresponds to the neutralino $\tilde Z_i$ having large $\tilde H^0_2$
component $(N_{i4})$. For the mixed region it corresponds to $\tilde
Z_2$, while the LSP $(\tilde Z_1)$ is dominantly a photino. But for
the higgsino dominated region both $\tilde Z_1$ and $\tilde Z_2$ have
large $\tilde H_2$ components. The large phase space available for $t
\rightarrow \tilde t_1 \tilde Z_1$ makes it the dominant decay mode
for this region, while $t \rightarrow \tilde t_1 \tilde Z_2$ is the
dominant one in the mixed region. Moreover, the overall SUSY branching
ratio ($B_S$) for top is significantly larger in the former case.
Consequently the higgsino dominated region is more vulnerable to the
constraints on $B_S$ from the top quark decay experiments
\cite{eleven,twelve}. 
\bigskip

\noindent {\large {\bf 3. \underbar{\bf RESULTS AND DISCUSSION} :}}
\medskip

\nobreak
It is well known that one cannot get as large a SUSY contribution to
$R_b ~(= .0045)$ as required by the central value of the data (3).  We
shall consider a contribution of about half this value, i.e.
\be
\delta R_b = .0018 - .0026,
\ee
as viable.  It would bring $R_b$ to within $1.6\sigma$ of the data
(i.e. the 90\% CL limit).  Moreover since $\Delta \alpha_s \simeq -
4\delta R_b$, it will exactly bridge the gap between $\alpha_s \simeq
0.124 \pm .007$ \cite{two} as measured from the $\Gamma_Z^{\rm had}$
and its global average value (2).  The upper limit on the SUSY
branching fraction of top decay is usually assumed to be
\be
B_S < 0.4,
\ee
from the top decay data \cite{seven,thirteen}.  The quantitative basis
of this assumption will be discussed later.

Although we shall make a detailed scan of the parameter space, it will
be useful to focus on three representative points in the $(M_2,\mu)$
plane, 
\be
A. ~(150,-40)GeV, ~~~ B. ~(60,-60)GeV, ~~~ C. ~(40,-70)GeV,
\ee
where $A$ belongs to the higgsino dominated region and $B,C$ to the
mixed region.  They have been chosen to give the most favourable
values of $\delta R_b$ in their respective regions, within the allowed
parameter space.  Table I shows the corresponding chargino and
neutralino masses and compositions for $\tan\beta = 1.1$, which is
close to its lower limit of $1$ \cite{one}.

For $A$, the $\tilde W_1$ mass of $70 ~GeV$ is very close to the
LEP-1.5 limit of $65 ~GeV$ \cite{fourteen}.  The $\tilde W_1$ is
higgsino dominated in both its $U$ and $V$ components, and so are the
two lightest neutralinos $\tilde Z_{1,2}$.  The former implies small
$Z\tilde W_1 \tilde W_1$ couplings (15) and hence a modest $\delta
R_b$ despite the low $\tilde W_1$ mass.  The latter implies large $t
\rightarrow \tilde t_1 \tilde Z_{1,2}$ branching ratio $(B_S)$, in
potential conflict with the top decay constraint (19).  For $B$ and
$C$, on the other hand, the charginos are roughly degenerate and
$\tilde W_2$ has a large gaugino (higgsino) component in $U(V)$.  This
implies large $Z\tilde W\tilde W$ couplings (15) and hence a more
favourable $\delta R_b$ despite the larger chargino mass.  Among the
lighter neutralinos only $\tilde Z_2$ has a large $\tilde H_2$
component, while the $\tilde Z_1$ is completely gaugino dominated
\cite{fifteen}.  This implies a comparatively smaller SUSY branching
ratio $(B_S)$ for top decay.  As we shall see below, the point $C$
gives far the best $R_b$ and least $B_S$, as desired.  However, it is
very close to the MSSM limit of
\be
M_2 > 36 ~GeV,
\ee
corresponding to a gluino mass limit of $m_{\tilde g} > 150 ~GeV$
\cite{sixteen} via eq. (8) \cite{seventeen}.  In other words $M_2
= 40 ~GeV$ implies a gluino mass $m_{\tilde g} = 160 ~GeV$ only.  Note
that the corresponding gluino mass for $M_2 = 60 ~GeV$ is $240 ~GeV$.
Thus both the chargino and gluino masses corresponding to the point
$B$ are safely above the reach of LEP-2 and Tevatron respectively.
What constrains this point is the LEP-1 limit, $M_{\tilde Z_1} +
M_{\tilde Z_2} \gsim M_Z$, which is not going to get any stronger.

The stop mass has the usual LEP bound \cite{two}, $m_{\tilde t_1} > 45
~GeV$.  There is also a constraint from the $D0\!\!\!/$ experiment
\cite{eighteen}, i.e.
\be
m_{\tilde t_1} \neq 65 - 88 ~GeV ~{\rm for}~ M_{\tilde Z_1} \leq 35
~GeV.
\ee
It is based on the neutral current decay mode
\be
\tilde t_1 \rightarrow c \tilde Z_1,
\ee
assuming $m_{\tilde t_1} < M_{\tilde W_1} + m_b$.  Thus it applies to
$B$ and $C$, but not for the higgsino dominated case $A$.
\bigskip

\noindent {\bf SUSY contributions to $R_b$ and top $BR$:}
\medskip

\nobreak
Fig. 1 shows the SUSY contributions to $R_b$ and the top $BR$ as
functions of the stop mixing angle $\theta_{\tilde t}$ and the lighter
stop mass $m_{\tilde t_1}$.  The three parts of the figure (a,b,c)
correspond to the three cases $A,B,C$ respectively.  The SUSY $R_b
~(\delta R_b)$ is clearly seen to peak at small negative value of
$\theta_{\tilde t} ~(\simeq -5^\circ)$ as expected from (13,14).  On
the other hand the SUSY $BR$ $(B_S)$ curves peak at large positive
values of $\theta_{\tilde t}$ as per (16,17).  Its insensitivity to
$\theta_{\tilde t}$ for the case $A$ is due to the fact that both the
higgsino dominated neutralinos are kinematically accessible in this
case to top quark decay.  The range $-15^\circ \leq \theta_{\tilde t}
\leq 0$ represents an optimal range for getting a large $\delta R_b$
along with a modest $B_S$.

The numerical values of these quantities show a striking difference
between the higgsino dominated case (A) and the mixed cases (B,C).  In
the former case (Fig. 1a) the SUSY $BR ~(B_S)$ is generally larger
than the Tevatron constraint (19).  One gets a $B_S$ of $.46$ for
$m_{\tilde t_1} \simeq 80 ~GeV$ and $\theta_{\tilde t} \simeq
-15^\circ$, corresponding to $\delta R_b = .0014$.  One can of course
suppress $B_S$ by going to a higher $m_{\tilde t_1}$ along with a lower 
$|\theta_{\tilde t}|$ \cite{nineteen}.  It is clear however that one cannot
get any significant enhancement of $\delta R_b$.  Thus the higgsino
dominated region cannot give a $\delta R_b$ in the required range of
(18) within the top decay constraint (19).  In contrast the mixed case
$B$ (Fig. 1b) gives $\delta R_b = .0018 - .0022$ with $B_S = 0.3 -
0.4$ for $m_{\tilde t_1} = 70 - 60 ~GeV$ and $\theta_{\tilde t} \simeq
-15^\circ$.   The mixed case $C$ (Fig. 1c) gives even a better $\delta
R_b = .0020 - .0024$ with $B_S = 0.3 - 0.4$ for $m_{\tilde t_1} = 70 -
50 ~GeV$ and $\theta_{\tilde t} \simeq -15^\circ$.  Thus in the mixed
region one can get a significant $\delta R_b$ in the range (18) within
the top decay constraint on $B_S$ (19).  Note that one could trade off
a lower value of $|\theta_{\tilde t}|$ for a higher stop mass without
changing $\delta R_b$ and $B_S$.  Similarly one can trade off a higher
$\tan\beta$ for a lower stop mass, as we shall see below.  This will
be useful for keeping the stop mass within the $D0\!\!\!/$ constraint
(22) \cite{eighteen}.

Fig. 2 shows the SUSY contributions to $R_b$ $(\delta R_b)$ and top $BR
(B_S)$ as contour plots in the $M_2,\mu$ plane for, $\theta_{\tilde t}
= -15^\circ$, $\tan\beta = 1.1$ and stop masses of $50$ and $60 ~GeV$,
which are below the $D0\!\!\!/$ excluded region (22).  The region
excluded by LEP-1 and LEP-1.5 $(M_{\tilde W_1} > 65 ~GeV)$ are
indicated.  One gets the best value of $\delta R_b$ close to the
boundary of this region as expected.  Much of the remainder is
excluded however by the condition $m_{\tilde t_1} > M_{\tilde Z_1}$.
One sees a steady increase of $\delta R_b$ and decrease of $B_S$ as
one moves down from the higgsino dominated region to the mixed one by
decreasing the ratio $M_2/|\mu|$.  The three points $A,B$ and $C$ of
(20) are indicated by dots.  One sees a 30\% increase in $\delta R_b$
along with a similar drop in $B_S$ as one moves down from the higgsino
dominated point $(A)$ to the mixed ones $(B,C)$.  By far the best
values of $\delta R_b$ and $B_S$ are obtained for the last point $C$;
but it is close to the gluino mass limit (21), represented by the
$x$-axis.  Finally one sees a 10\% (20\%) drop in $\delta R_b$ $(B_S)$
by increasing the stop mass from $50$ to $60 ~GeV$.

Fig. 3 shows the analogous contour plots of $\delta R_b$ and $B_S$ for
$\tan\beta = 1.4$.  In going from $\tan\beta = 1.1$ to $1.4$ one sees
only a 15\% drop in $\delta R_b$ and $B_S$.  This is because the
decrease of $1/\sin^2\beta$ in (13,14) and (16,17) are partly offset
by the drop in the $\tilde W_1$ and $\tilde Z_{1,2}$ masses.
Increasing $\tan\beta$ to $1.6$ would result in a further drop of only
5\% in these quantities.  However the drop in $\tilde Z_{1,2}$ mass
brings these points right on to the LEP boundary line.

Table II lists the values of $\delta R_b$ and $B_S$ for the mixed
cases $B$ and $C$ of Figs. 2 and 3.  The values lying within the range
of (18) and (19) are ticked as viable solutions.  Most of the viable
solutions correspond to the case $C$.  Note however that there is one
viable solution for the point $B$ with $\tan\beta = 1.4$ and a stop
mass of $60 ~GeV$.  It is an important point as it is not very close
to the lower limits of the relevant SUSY masses and $\tan\beta$.  This
is essentially the same as the solution advocated in \cite{eight}.

Table II also shows that the point $C$ can give viable solutions for
stop masses of $90 - 100 ~GeV$, which lie above the $D0\!\!\!/$
excluded region (22).  This is an interesting case, where one expects
charged current decay of stop,
\be
\tilde t_1 \rightarrow b\tilde W_1, ~\tilde W_1 \rightarrow \tilde Z_1
\ell \nu(qq').
\ee
Its phenomenological implications will be discussed at the end of this
section. 
\bigskip

\noindent {\bf Impact on Top Quark Phenomenology:}
\medskip

\nobreak
We shall discuss the phenomenological impact of SUSY decay on the
Tevatron top quark events, concentrating on the stop mass range of $50
- 60 ~GeV$.  In this case the viable solutions to $\delta R_b$
correspond to a 
\be
B_S = 0.3 - 0.4.
\ee
The most important sample of top events comes from the isolated lepton
plus multijet events with at least $1$ $b$-tag, which satisfy a lepton
and a missing  $E_T$ cut of $E^\ell_T > 20 ~GeV$ and $E\!\!\!\!/_T > 20
~GeV$ \cite{eleven,twentyone}.  For the SUSY contribution, one of the
top quarks decays into
\be
t \rightarrow \tilde t_1 \tilde Z_i, ~\tilde t_1 \rightarrow c \tilde
Z_1,
\ee
while the other undergoes SM decay
\be
\bar t \rightarrow \bar b W, ~W \rightarrow \ell \nu.
\ee
The dominant $\tilde Z_i$ in the SUSY decay is $\tilde Z_1 ~(\tilde
Z_2)$ for the higgsino dominated (mixed) region.  Thus the total
number of events will be suppressed by a factor
\be
(1 - B_S)^2 + B_S(1 - B_S) {3 \over 4} \cdot {\epsilon (E\!\!\!\!/_T)
\over 1 - \epsilon_b/2},
\ee
where the two terms represent the SM and SUSY contributions.  Here
$\epsilon_b$ is the $b$-tagging efficiency and $\epsilon
(E\!\!\!\!/_T)$ represents the efficiency of satisfying the
$E\!\!\!\!/_T > 20 ~GeV$ cut for the SUSY contribution relative to the
SM.  Substituting the experimental value for $\epsilon_b \simeq 0.24$
\cite{eleven,twentyone} with our estimate of $\epsilon (E\!\!\!\!/_T)
\simeq 1.06$, the above factor can be approximated by
\be
(1 - B_S)^2 + B_S (1 - B_S).
\ee
Thus the fraction of SUSY contribution to these $t\bar t$ events is
$\simeq B_S$.

In order to proceed further we have to consider the distribution in
the number of jets $(\sigma_n)$.  As per the CDF jet algorithm the
$E^{\rm jet}_T$ is obtained by combining all the hadronic $E_T$ within
an angular radius of $0.7$ in the $\eta,\phi$ plane; and all the jets
with $E^{\rm jet}_T > 15 ~GeV$ coming within the rapidity range
$|\eta_{\rm jet}| < 2$ are counted.  We shall follow a poor man's
prescription of incorporating the effects of hadronisation and QCD
radiation in a parton level Monte Carlo program by increasing the
$E^{\rm jet}_T$ threshold from $15$ to $20~GeV$ and transferring 30\%
of $\sigma_n$ to $\sigma_{n+1}$ \cite{twentytwo}.  This prescription
seems to give reasonable agreement with ISAJET results.  It should be
adequate for our purpose, which is to estimate the difference between
the $\sigma_n$ distributions of the SM and the SUSY contributions.

Table III shows the fractional $\sigma_n$ distribution of the SM along
with the SUSY contributions for the mixed and the higgsino dominated
cases.  The SUSY contributions are seen to favour fewer number of jets
compared to the SM.  This is very pronounced for the higgsino
dominated region due to the large $t \rightarrow \tilde t_1 \tilde
Z_1$ contribution.  But even for the mixed region of our interest
there is a clear preference for fewer jets compared to the SM.  This
has several implications for top quark phenomenology, as we see below.

\begin{enumerate}
\item[{i)}] Compared to the SM expectation of 10\% of the $t\bar t$
events in the 2-jet channel, one expects an additional contribution of
\be
(.35 - .10) B_S = .25 B_S,
\ee
i.e. $7.5$ to $10\%$ using (25).  The 4th column shows $6.4$ expected
$t\bar t$ events in the SM from the CDF MC \cite{eleven}, which we
expect to include $1-2$ from the $\ell \ell$ and $\ell \tau$ channels
of $t\bar t$ decay.  Correspondingly we expect $\sim 4$ more 2-jet
events from $(25 - 30)$.  This will evidently be favoured by the
central value of the data shown in the last column, though the errors
are too large to draw any definite conclusion.  Similarly one expects
a deficit of $\sim 4$ events in the $\geq 4$ jet, which is also
compatible with data. 

\item[{ii)}] The CDF $\bar tt$ cross-section is based on the sample of
$\geq 3$ jet events.  Correspondingly the suppression factor (29)
becomes
\be
(1 - B_S)^2 + \epsilon_3 B_S(1 - B_S),
\ee
where $\epsilon_3$ represents the efficiency of surviving the $\geq 3$
jet cut for the SUSY contribution relative to the SM.  We see from
Table III that $\epsilon_3 = 2/3 ~(1/4)$ for the mixed (higgsino
dominated) region.  Thus the mixed region of our interest corresponds
to a suppression factor of 
\be
(1 - B_S) (1 - B_S + 2 B_S/3) = 2/3 - 1/2,
\ee
for $B_S = 0.3 - 0.4$.  Therefore the CDF $\bar t t$ cross-section
should be compatible with $2/3$ to $1/2$ times the QCD value.  From
the CDF data \cite{eleven},
\be
\sigma_t = 7.5 \pm 1.8 ~pb, ~m_t = 175.6 \pm 9 ~GeV,
\ee
it is evident that the central value of their cross-section is already
higher than QCD estimate of $\sigma_t (175) = 5.5 ~pb$
\cite{twentythree}.  However taking a $1.64 \sigma ~(90\% ~{\rm CL})$
lower limit on both the quantities would correspond to a $\sigma_t$ of
$4.5 ~pb$ to be compared with a QCD estimate of $\sigma_t (160) = 9
~pb$ \cite{twentythree}.  Thus a SUSY BR of $0.4$ and $\epsilon_3 = 2/3$
is barely compatible
with the CDF data \cite{twentyfour}.  The corresponding compatibility
with the $D0\!\!\!/$ cross-section \cite{twelve},
\be
\sigma_t = 5.3 \pm 1.6 ~pb,
\ee
is evidently easier to satisfy.  It may be noted here that for the
higgsino dominated region, the range of $B_S > 0.45$ and $\epsilon_3 = 1/4$
would
correspond to a suppression factor $< 1/3$, in clear conflict with the
CDF data.

\item[{iii)}] The SUSY contribution to the sample of $\geq 3$ jet
events accounts for a fraction
\be
2B_S/(3-B_S) = .22 - .31.
\ee
This means a $20 - 30\%$ drop in the number of $t\bar t$ events in the
dilepton as well as the double $b$-tag events compared to the SM
prediction if the $t \bar t$ cross-section is normalised to the $b$-tagged
$\ell ~+ \geq 3~jet$ sample.  The present CDF data seems to have 
$\sim 8$ events of
each kind, of which the $20-30\%$ drop is within a $1\sigma$ effect.

\item[{iv)}] The SUSY contribution has several kinematic
distributions, which are distinct from the SM.  Fig. 4 shows the
transverse mass distribution of the lepton and the missing $E_T$
\be
M_T = 2E^\ell_T ~E\!\!\!\!/_T (1 - \cos\Delta\phi),
\ee
where the SUSY contribution shows a clear tail beyond the Jacobian
peak of the SM contribution.  The SM contribution shown corresponds to
the $\geq 3$ jet sample of the CDF $t\bar t$ Monte Carlo including
hadronisation and detector resolution effects \cite{twentyfive}, which
are responsible for the spillover to the $M_T > M_W$ region.  In
contrast the SUSY contribution corresponds to our parton level MC
without these effects, which gives only a conservative estimate of its
tail.  Nonetheless 30\% of the SUSY contribution is seen to occur at
$M_T > 120 ~GeV$, in contrast to a 10\% spillover for the SM.  This
corresponds to an excess of 
\be
.2 \times 2 B_S/(3 - B_S) = .044 - .062,
\ee
i.e. an excess over the SM prediction by about 50\%.  The present data
sample of CDF corresponds to a SM prediction of $\sim 4$ events with
$M_T > 120 ~GeV$, of which the 50\% excess constitutes a $1\sigma$
effect.  With an order of magnitude increase in the data sample
following the main injector run one expects an excess of at least
$3\sigma$.  Similarly the predicted deficit of $20-30\%$ in the
dilepton and double $b$-tag events will each constitute at least a
$2\sigma$ effect.  These will provide important tests of the SUSY
contribution, since one expects very little background in each case.
\end{enumerate}
\bigskip

\noindent {\bf The $90-100~GeV$ Stop Case:}
\medskip

\nobreak
Finally we consider the phenomenological implications of a stop mass
of $90 - 100~GeV$, which was shown to give a viable contribution to
$R_b$ for the case $C$.  This is an interesting scenario since it
corresponds to a modest $B_S$ of $\sim 0.2$.  Besides the stop can
undergo charged current decay (24) leading to a soft but visible
lepton.  The resulting efficiency for the $E^\ell_T > 20 ~GeV$ cut is
$\sim 1/2$ that of the SM decay.  Consequently the overall lepton
detection efficiency of the SUSY contribution is $\sim 3/2$ higher
than the previous case.  This also implies a somewhat larger
efficiency $\epsilon_3$ for the $\geq 3$ jet cut compared to the
previous case.  Taking account of these efficiency factors leads to an
overall suppression factor of
\be
1 - B_S \simeq 0.8,
\ee
which is modest compared to (32).  Thus the CDF top quark data can
accommodate this case more easily.

The most interesting phenomenological test of this scenario follows from
the pair production of stops, followed by their leptonic decay (24).
This leads to a signal of isolated but relatively soft dilepton events
\cite{ten}.  We have estimated the resulting dilepton signal using our
parton level MC with
\[
|\eta_\ell| < 1, ~20^\circ < \phi_{\ell^+\ell^-} < 160^\circ,
~E\!\!\!\!/_T > 20 ~GeV
\]
and
\be
E^\ell_T > 10 (15) ~GeV.
\ee
We estimate a signal cross-section of $100 (80) ~fb$ for a stop mass
of $90~GeV$.  It should be noted here that this signal has been
recently analysed in \cite{twentysix} using the ISAJET program.  With
the above cuts they estimate a dilepton signal of similar magnitude,
which also has an accompanying jet of $E_T > 15 ~GeV$.  This jet helps
to control the $W^+W^-$ background.  Moreover they have used a cut on
the scalar sum
\be
|E^{\ell^+}_T| + |E^{\ell^-}_T| + |E\!\!\!\!/_T| < 100 ~GeV
\ee
to suppress the $t\bar t$ as well as the $W^+W^-$ background without
affecting the signal seriously.  Thus with the current CDF luminosity
of $0.11 ~fb^{-1}$ one expects $\sim 10$ soft but isolated 
dilepton events for a
$90~GeV$ stop undergoing charged current decay.  There is only a
modest drop in the signal rate for a stop mass of $100~GeV$.
\bigskip

\noindent {\large {\bf 4. SUMMARY}}
\medskip

\nobreak
The SUSY contributions to $R_b$ and top quark decay are studied
simultaneously over the relevant MSSM parameter space to obtain an
optimal solution to the $R_b$ anomaly within the constraint of the top
quark data.  Contrary to the popular notion the higgsino dominated
region $(|\mu| \ll M_2)$ is disfavoured on both counts.  It makes a
relatively small contribution to $R_b$ $(\delta R_b)$ along with an
excessively large one to top $BR ~(B_S)$.  On the other hand one gets
a 30\% increase in $\delta R_b$ along with a similar drop in $B_S$ by
going to the mixed region $(|\mu| \sim M_2)$, which corresponds to a
photino dominated LSP.  We have focussed on two points belonging to
this region -- i.e. $M_2,\mu = 60,-60~GeV$ and $40,-70~GeV$.  The
latter offers by far the best values of $\delta R_b$ and $B_S$.  But
it is close to the boundary of the region disallowed by the Tevatron
limit on gluino mass, while the former lies safely above the reaches
of Tevatron as well as LEP-2.  Both give acceptable solutions to the
$R_b$ anomaly for stop mass of $50-60~GeV$.  We analyse the
corresponding predictions for top quark decay, which can be tested
with Tevatron data.  The latter point also gives acceptable solution
for a stop mass range of $90 - 100~GeV$.  In this case one expects a
distinctive dilepton signal from the pair production of stop followed
by its charged current decay, which can again be tested with the
Tevatron data.

\bigskip

\noindent {\large {\bf Acknowledgements}}
\medskip

\nobreak
This investigation was started as a working group project in WHEPP-IV,
Calcutta, last January.  We thank the organisers of this workshop for
their kind hospitality and Michelangelo Mangano for his participation
in the initial stages of this work.  The work of MD was supported in 
part by the US Department of Energy under grant no. DE-FG02-95ER40896,
by the Wisconsin Research Committee with funds granted by the 
Wisconsin Alumni Research Foundation, as well as by a grant from the 
Deutsche Forschungsgemeinschaft under the Heisenberg program. 
The work of RMG is partially supported by a
grant (No: 3(745)/94/EMR(II)) of the Council of Scientific and
Industrial Research, Government of India, while that of SR is
partially funded by a project (DO No: SERC/SY/P-08/92) of the
Department of Science and Technology, Government of India.

\newpage

\begin{center}
{\large \underbar{Figure Captions}}
\end{center}
\bigskip\bigskip

\begin{enumerate}
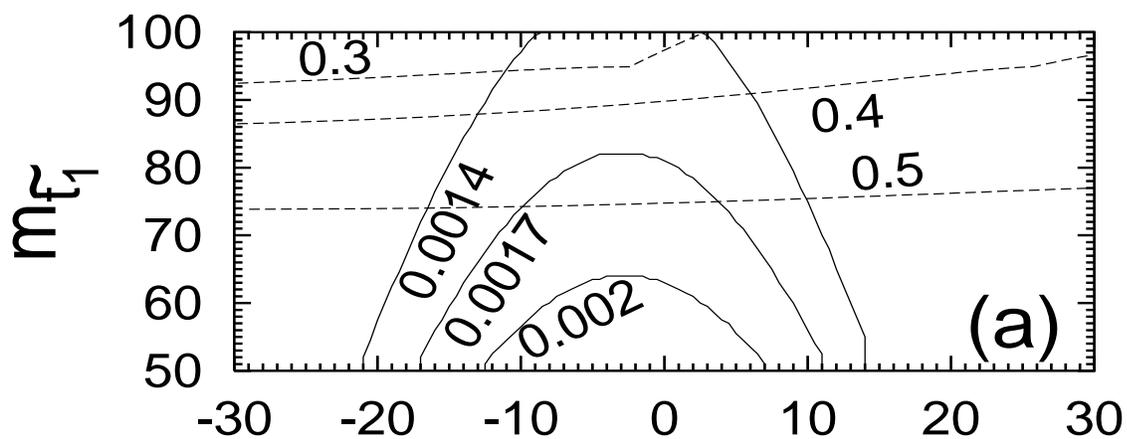
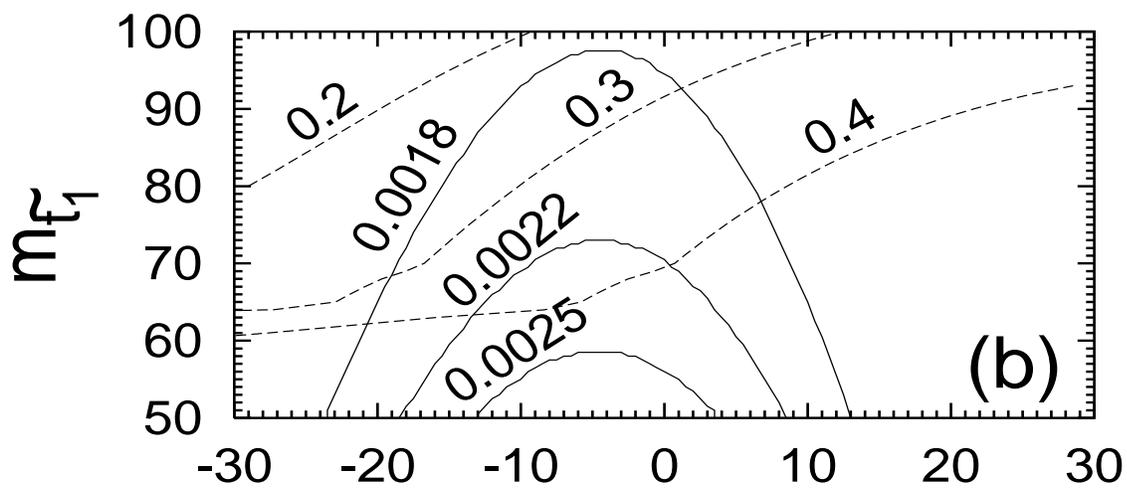
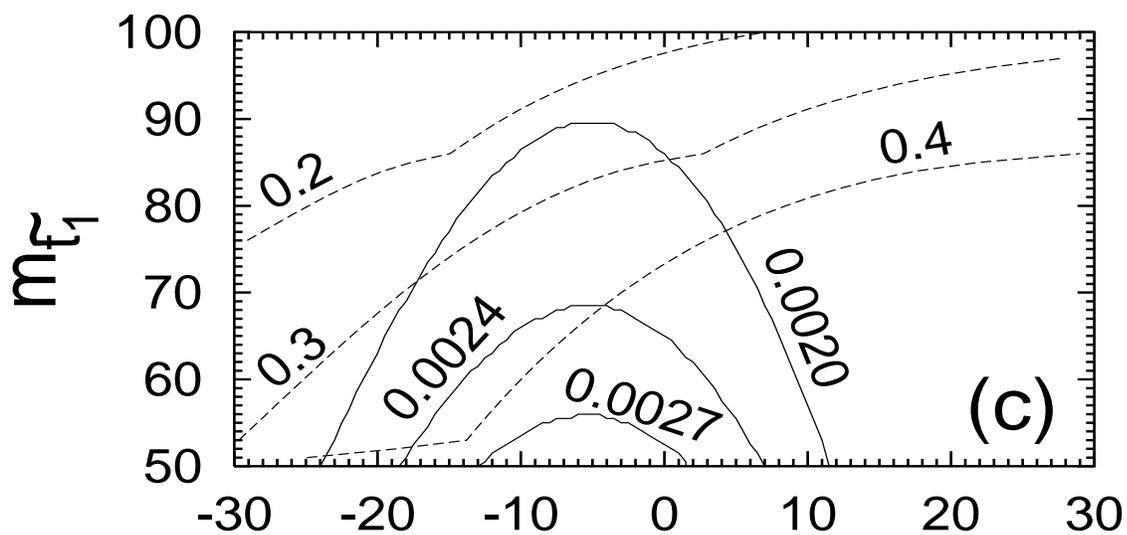
\item[{Fig.~1.}] SUSY contributions to $R_b$ (solid) and the top $BR$
(dashed) are shown as contour plots in stop mass and mixing angle for
$M_2,\mu = {\rm (a)} ~150,-40 ~{\rm (b)}~ 60,-60 ~{\rm (c)}~
40,-70~GeV$, with $\tan\beta = 1.1$.

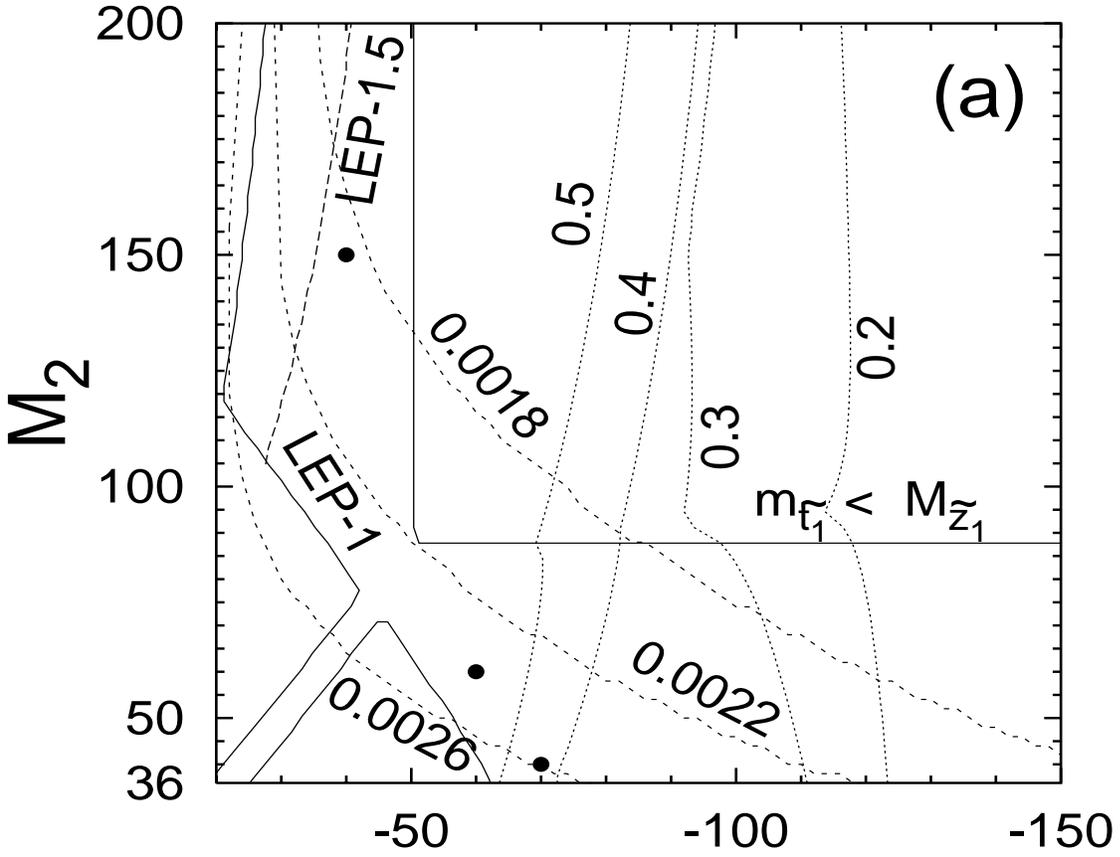
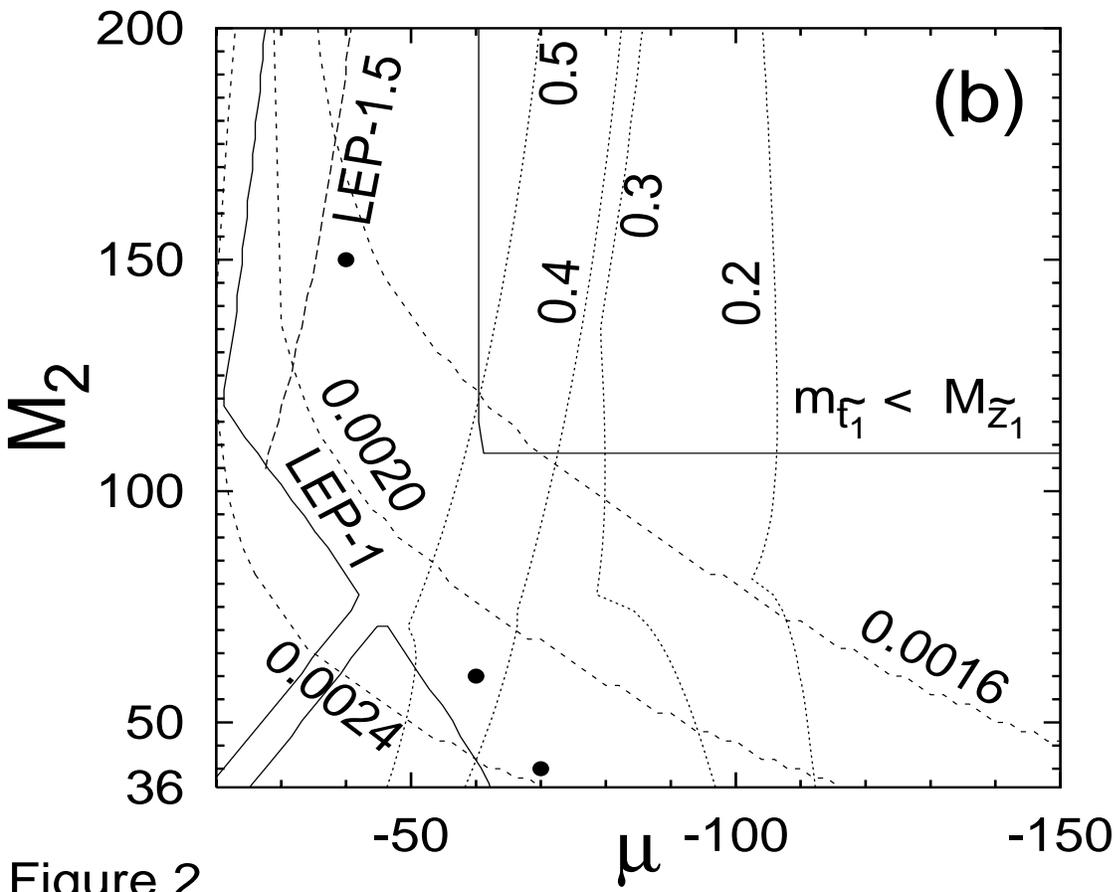
\item[{Fig.~2.}] SUSY contributions to $R_b$ (dashed) and top $BR$
(dotted) are shown as contour plots in the $M_2,\mu$ plane for stop
mass of (a) $50$ and (b) $60~GeV$ with $\theta_{\tilde t} = -15^\circ$
and $\tan\beta = 1.1$.  The $x$-axis correspnds to the boundary of the
region disallowed by the Tevatron limit $(m_{\tilde g} > 150~GeV)$.
Bullets show the parameter choices A,B,C of equation (20).

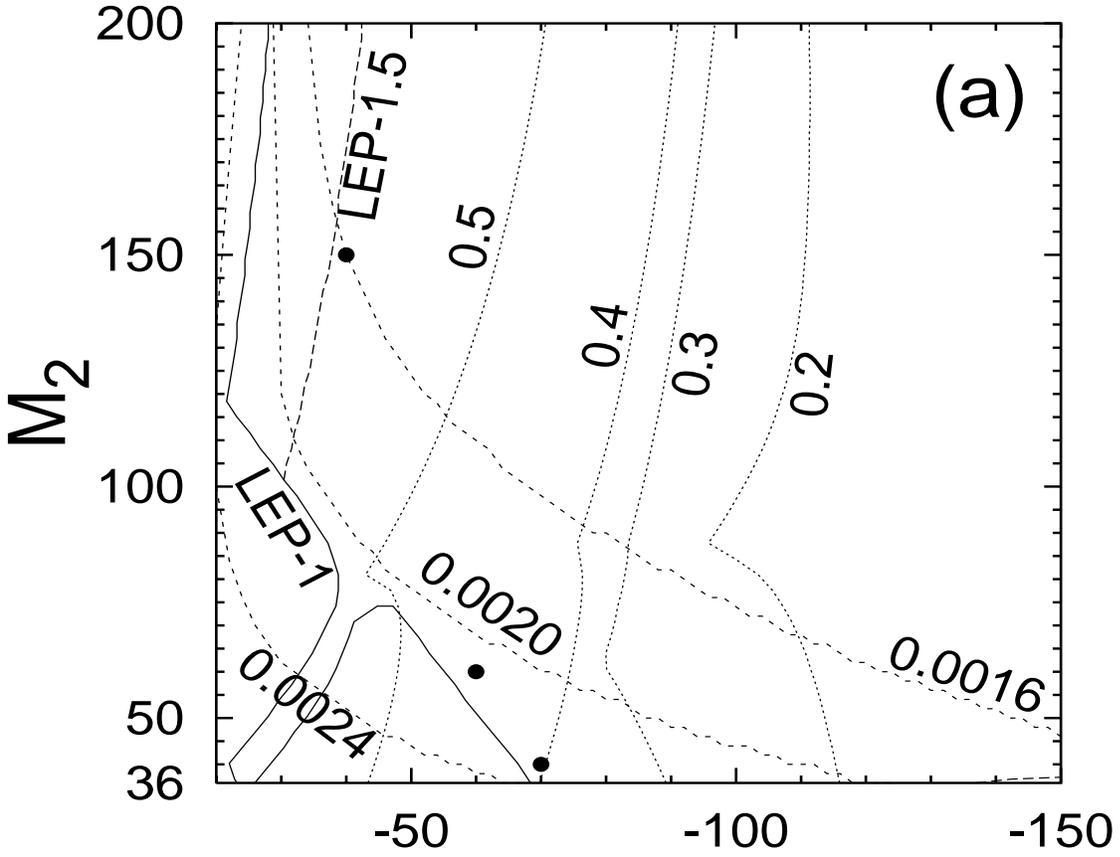
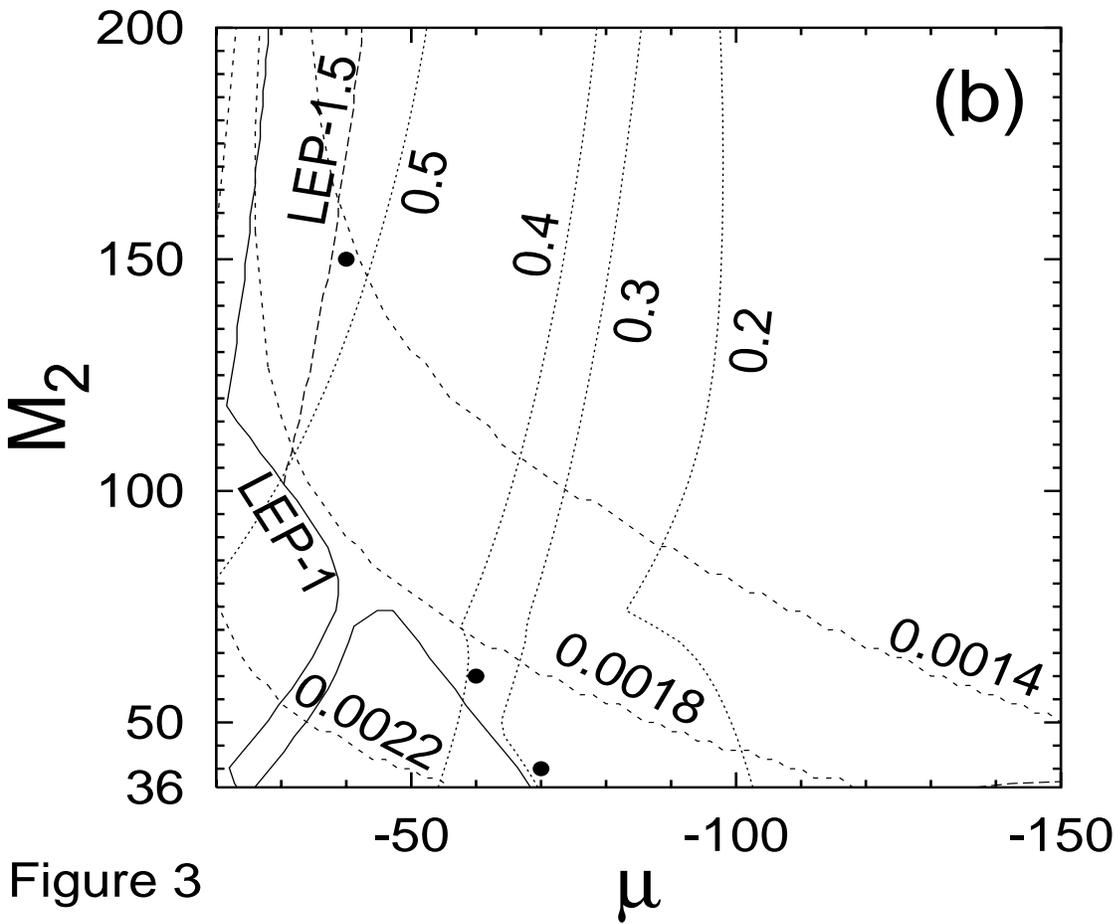
\item[{Fig.~3.}] Same as Fig. 2 for $\tan\beta = 1.4$.  The boundary
of the region $m_{\tilde t_1} < M_{\tilde Z_1}$ is not shown to avoid
overcrowding.

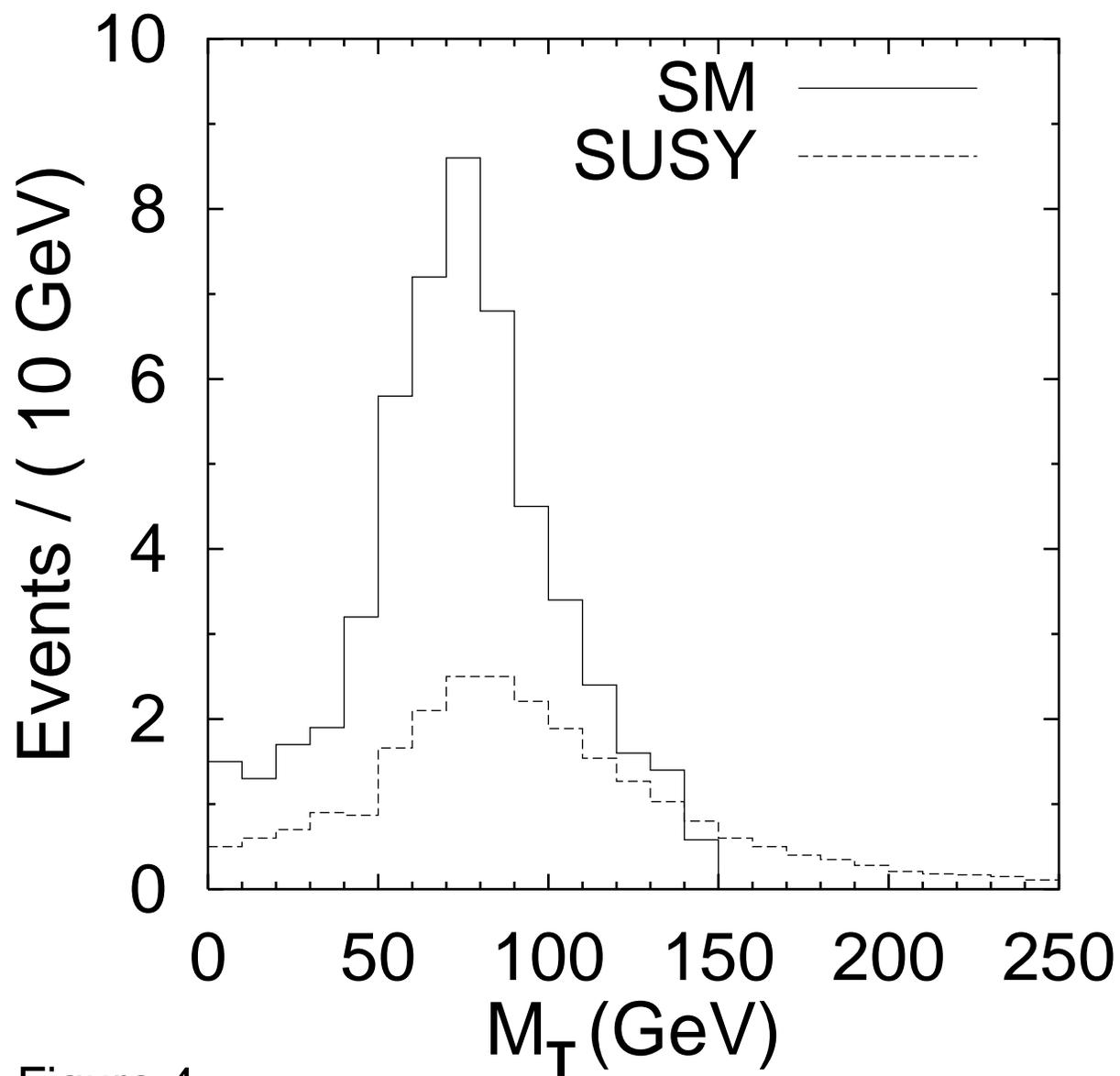
\item[{Fig.~4.}] The $(\ell, E\!\!\!\!/_T)$ transverse mass
distribution of SM and SUSY contributions to the $t\bar t$ events with
free normalisation. The former is taken from the CDF MC of \cite{twentyfive}
including hadronisation and detector resolution, while the latter is
based on our parton level $MC$ result without these effects.
\end{enumerate}

\newpage

\begin{table}
{\protect\footnotesize Table I : Chargino and neutralino masses and
compositions for three representative points in the $M_2, \mu$
plane, corresponding to optimal values of $R_b$ ($\tan \beta$ = 1.1)}.

\bigskip
\begin{center}
\begin{tabular}{|ccccc|} \hline
$M_2 , \mu$ (GeV) & $M_{\tilde W_i}$ (GeV) & $U_{ij} / V_{ij}$ &
$M_{\tilde Z_i}$ (GeV) & $N_{ij}$ \\ \hline
& 70 & - .31, .95 & 40 & - .02, .01, .72, .68 \\
&& .37, - .92 & 80 & - .25, .31, - .63, .65 \\
150, -40 & 180 & .95, .31 & 85 & - .95, - .23, .11, - .15 \\
&& .92, .37 & 181 & - .15, .92, .23, - .27 \\ \hline
& 97 & .31, .95 & 36 & - .92, - .37, .09, - .07 \\
&& .95, - .31 & 60 & 0, .07, .74, .67 \\
60, -60 & 105 & .95, - .31 & 106 & - .28, .83, .28, - .39 \\
&& .31, .95 & 111 & - .27, .41, - .61, .63 \\ \hline
& 82 & .82, .57 & 24 & - .91, - .40, .06, - .05 \\
&& .93, .35 & 69 & - .05, .15, .76, .62 \\
40, - 70 & 113 & .57, - .82 & 89 & - .31, .79, .21, - .47 \\
&& - .35, .93 & 123 & - .26, .42, - .60, .62 \\ \hline
\end{tabular}
\end{center}
\end{table}

\newpage

\begin{table}
{\protect\footnotesize Table II : SUSY contributions to $R_b (\delta
R_b)$ and top BR ($B_s$) for the mixed region. The cases satisfying
(18) and (19) are ticked as viable solutions to the $R_b$ anomaly.}

\bigskip
\begin{center}
\begin{tabular}{|cccccc|} \hline
$M_2, \mu$&$\tan \beta$&$m_{\tilde t_1}$&$\delta R_b$&$B_s$& Remark \\ 
(GeV) && (GeV) &&& \\ \hline
60, -60 & 1.1 & 60 & .0022 & .45 & \\
($\theta_{\tilde t} = -15^\circ$) & 1.1 & 50 & .0024 & .53 & \\
& 1.4 & 50 & .0021 & .47 & \\
& 1.4 & 60 & .0019 & .40 & $\surd$ \\ \hline
40, -70 & 1.1 & 60 & .0024 & .37 & $\surd$ \\
$(\theta_{\tilde t} = -15^\circ)$ & 1.1 & 50 & .0026 & .46 & \\
& 1.4 & 50 & .0023 & .41 & $\surd$ \\
& 1.4 & 60 & .0021 & .30 & $\surd$ \\
& 1.1 & 90 & .0018 & .18 & $\surd$ \\
&&&&& \\
$(\theta_{\tilde t} = - 5^\circ)$ & 1.1 & 90 & .0020 & .23 & $\surd$ \\
& 1.1 & 100 & .0018 & .16 & $\surd$ \\ \hline
\end{tabular}
\end{center}
\end{table}

\newpage

\begin{table}
{\protect\footnotesize Table III : Fractional distribution of the SM
and SUSY contributions to the $t \bar t$ events in the number of jets.
The 4th column shows the numbers of expected $t \bar t$ events in the
SM from \cite{eleven}, while the 5th column shows the corresponding numbers
of observed (background) events.}

\bigskip
\begin{center}
\begin{tabular}{|c|ccc|cc|} \hline
$\sigma_n/\sigma$ & SM & SUSY & SUSY &
\multicolumn{2}{c|}{No. of CDF events} \\
$n$ && (mixed) & (higgsino) & $t \bar t$& observed (bg) \\ \hline
1 & --- & .05 & .25 & 0.8 & 70 $\pm$ 9 (69 $\pm$ 11) \\
2 & .10 & .35 & .50 & 6.4 & 45 $\pm$ 7 (28 $\pm$ 4) \\
3 & .40 & .40 & .25 & 12.8 & 18 $\pm$ 4 (6.5 $\pm$ 1) \\
$\geq$ 4 & .50 & .20 & --- & 16.7 & 16 $\pm$ 4 (2.6 $\pm$ .5) \\ \hline
\end{tabular}
\end{center}
\end{table}

\newpage

\begin{thebibliography}{99}

\bibitem{one} For a review see H. Haber and G.L. Kane, Phys. Rep. 117,
75 (1985). 
\bibitem{two} Review of Particle Properties, Phys. Rev. D50, 1173-1826
(1994). 
\bibitem{three} G. Bhattacharyya, G. Branco and W.S. Hou, NTUTH-95-11
(hep-ph/9512239); E. Ma and D. Ng, hep-ph/9505268; E. Ma,
Phys. Rev. D53, 2276 (1996); C.V. Chang, D. Chang and W.Y. Keung,
NHCU-HEP-96-1 (hep-ph/9601326); I. Montvay, DESY 96-047.
\bibitem{four} P. Chiapetta, J. Leyssac, F.M. Renard and
C. Verzegnassi, hep-ph/9601306; G. Altarelli, N. Di Bartolomeo,
F. Feruglio, R. Gatto and M. Mangano, hep-ph/9601324; V. Barger,
K. Cheung and P. Langacker, MADPH-96-936; K. Agashe, M. Graesser,
I. Hinchliffe and M. Suzuki, LBL-38569 (1996).
\bibitem{five} LEP Electroweak Working Group Report, LEPEWWG/96-01 (1996).
\bibitem{six} M. Boulware and D. Finnell, Phys. Rev. D44, 2054 (1991).
The last equation in this paper has a typographic error; the $C_{11}$
should be $C_{12}$.
\bibitem{seven} J.D. Wells, C. Kolda and G.L. Kane, Phys. Lett. B338,
219 (1994); D. Garcia, R. Jimenez and J. Sola, Phys. Lett. B347, 321
(1995); P.H. Chankowski and S. Pokorski, Phys. Lett. B356, 307 (1995);
J. Wells and G.L. Kane, Phys. Rev. Lett. 76, 869 (1996); J. Ellis,
J. Lopez and D. Nanopoulos, hep-ph/9512288; E. Ma and D. Ng,
hep-ph/9508338; A. Brignole, F. Feruglio and F. Zwirner, hep-ph/9601293.
\bibitem{eight} P.H. Chankowski and S. Pokorski, IFT-96/6
(hep-ph/9603310). 
\bibitem{nine} G. Passarino and M. Veltman, Nucl. Phys. B160, 151
(1979). 
\bibitem{ten} H. Baer, M. Drees, R.M. Godbole, J.F. Gunion and
X. Tata, Phys. Rev. D44, 725 (1991).
\bibitem{eleven} CDF Collaboration: A. Carner, Rencontres de Physique
de la Vallee d'Aoste, La Thuile, Italy (1996).  J. Huston, Pheno96
Symposium, Madison, Wisconsin, April 1996.
\bibitem{twelve} $D0\!\!\!/$ Collaboration: H. Schellman, Pheno96
Symposium, Madison, Wisconsin, April 1996.
\bibitem{thirteen} S. Mrenna, C.P. Yuan, Phys. Lett. B367, 188 (1996).
These authors favour a stronger bound of $B_S < 0.25$.
\bibitem{fourteen} ALEPH Collaboration: CERN-PPE/96-10; OPAL
Collaboration: CERN-PPE/96-019 and 020.
\bibitem{fifteen} One can easily check that in both cases the photino
component of $\tilde Z_1$ is 99\%.
\bibitem{sixteen} $D0\!\!\!/$ Collaboration: S. Abachi et. al.,
Phys. Rev. Lett. 75, 618 (1995); CDF Collaboration: J. Hauser,
Proc. 10th Topical Workshop on Proton-Antiproton Collider Physics,
Fermilab, May 1995.
\bibitem{seventeen} The physical gluino mass is related to the running
mass $M_3 (8)$ via the QCD correction factor, i.e. $m_{\tilde g} = M_3
[1 + 4.2 \alpha_S/\pi]$.  See e.g. N.V. Krasnikov, Phys. Lett. B345,
25 (1995); S.P. Martin and M.T. Vaughn, Phys. Lett. B318, 331 (1993).
\bibitem{eighteen} $D0\!\!\!/$ Collaboration: S. Abachi et. al.,
Fermilab-pub-95-380-E (Submitted to Phys. Rev. Lett.).
\bibitem{nineteen}  In this case one also expects a relaxation of the
$B_S$ limit (19) because the stop will undergo charged current decay
\cite{twenty}. 
\bibitem{twenty} J. Sender, hep-ph/9602354.
\bibitem{twentyone} CDF Collaboration: F. Abe et. al.,
Phys. Rev. Lett. 74, 2626 (1995).
\bibitem{twentytwo} M. Mangano (private communication).
\bibitem{twentythree} E.L. Berger, H. Contopanagos, ANL-HEP-PR-95-85,
hep-ph/9512212; S. Catani, M. Mangano, P. Nason, L. Trentadue,
CERN-TH/96-21, hep-ph/9602208.
\bibitem{twentyfour} In practice it suffices to reduce $m_t$ to $165
~GeV$ when the corresponding reduction in $B_S$ is taken into account.
The drop in $\delta R_b$ is offset by the rise in $R^{SM}_b$.
\bibitem{twentyfive} CDF Collaboration: F. Abe et. al.,
Phys. Rev. D50, 2966 (1994).
\bibitem{twentysix} H. Baer, J. Sender and X. Tata, Phys. Rev. D50,
4517 (1994).
\end{thebibliography}
\end{document}